\documentclass[11pt]{article}
\pdfoutput=1

\usepackage[usenames,dvipsnames]{xcolor}
\usepackage[a4paper]{geometry}
\usepackage[english]{babel}
\usepackage{cite,enumerate,enumitem,booktabs,float,graphicx}

\usepackage[T1]{fontenc}
\usepackage[utf8]{inputenc}
\usepackage{lmodern}

\makeatletter
\g@addto@macro\bfseries{\boldmath}
\makeatother

\usepackage{bm,amsmath,amssymb,tensor,mathtools}
\numberwithin{equation}{section}

\usepackage{mathrsfs}		
\usepackage[scr=rsfs]{mathalfa}

\usepackage{feynmp-auto}

\usepackage{caption}
\usepackage{subcaption}

\usepackage[pdftex]{hyperref}
\definecolor{dark-blue}{rgb}{0.15,0.15,0.4}
\hypersetup{
  colorlinks,
  linkcolor={dark-blue},
  citecolor={blue}
}

\newcommand{\beq}{\begin{equation}}
\newcommand{\eeq}{\end{equation}}

\renewcommand{\hat}{\widehat}

\newcommand{\normord}[1]{\vcentcolon\mathrel{#1}\vcentcolon}

\renewcommand{\SS}{\mathcal{S}}

\newcommand{\bra}{ \langle {\rm out} | }
\newcommand{\ket}{ | {\rm in} \rangle }

\usepackage{authblk}

\title{\textbf{%
	Soft zero for cylindrical gravitational waves
  }
}
\date{}

\author[]{Robert Penna\footnote{pennar@sunypoly.edu}}
\affil[]{%
  Department of Mathematics and Physics,
  \protect\\
  SUNY Polytechnic Institute,
  Utica, NY 13502 USA
}

\begin{document}
\unitlength = 1mm 
\maketitle
\thispagestyle{empty}

\begin{abstract}

The graviton $\SS$-matrix has a famous soft pole.  
We show that the $\SS$-matrix for cylindrical gravitational waves has a soft zero.  
The soft pole for ordinary gravitons comes from a Ward identity for supertranslation symmetry at asymptotic infinity.  
We show that the soft zero for cylindrical gravitational waves comes from a Ward identity for Geroch symmetry at asymptotic infinity.  
Because it is a zero rather than a pole, there is no memory effect.  
Overall, this soft zero is a manifestation of Geroch symmetry and of the extraordinary simplicity of cylindrical gravitational waves.

\end{abstract}

\newpage


\section{Introduction}

In recent years, there has been a lot of interest in ``infrared triangles'' \cite{Strominger:2017zoo}.  
An infrared triangle is a triangle of ideas relating asymptotic symmetries, soft theorems, and memory effects.  
Part of the interest in infrared triangles comes from the fact that they could be a stepping stone toward understanding holography for asymptotically flat gravity.  
The present paper came from trying to understand if there is an infrared triangle for cylindrical gravitational waves.  
Cylindrical gravitational waves enjoy a somewhat mysterious infinite dimensional symmetry called Geroch symmetry \cite{Geroch:1970nt,Geroch:1972yt,Breitenlohner:1986um,Nicolai:1991tt,Schwarz:1995td,Schwarz:1995zy,Lu:2007zv}.  
We wanted to understand if Geroch symmetry fits into an infrared triangle.

Cylindrical gravitational waves are effectively two dimensional.  
Our approach will be to dimensionally reduce general relativity to get a nonlinear sigma model and then to quantize the sigma model.  A more physically realistic approach would be to quantize gravity first and then dimensionally reduce\footnote{An interesting middle path is to dimensionally reduce the classical theory to three spacetime dimensions \cite{Ashtekar:1996cd,Bosma:2023sxn}.}.  However, we are going to confine our study to tree level physics.  Tree level physics is essentially classical physics and the sigma model describes classical cylindrical gravitational waves perfectly.  

Our main result will be to show that cylindrical gravitational waves have a soft zero.  This stands in stark contrast to the famous soft pole of ordinary gravitons.   
In a recent work, we computed some cylindrical gravitational wave scattering amplitudes directly using Feynman diagrams and the soft zero is already visible in those amplitudes \cite{penna2024a}.  
The main contribution of the present work is to provide a symmetry explanation for this soft zero.  
This zero is similar to an Adler zero because the dimensionally reduced theory is a nonlinear sigma model.  However, it is not exactly an ordinary Adler zero because one of the symmetries we consider (eq. \ref{eq:shift1}) is not exactly a shift symmetry and because the theory does not have $r$-translation invariance.

We will formulate the soft zero as a Ward identity for Geroch symmetry at asymptotic infinity. This is a piece of the infrared triangle for cylindrical gravitational waves.  The original examples of infrared triangles were  for gauge theories and asymptotically flat gravity\footnote{There has also been interesting work on the relationship between soft theorems and  symmetries for scalar fields \cite{Low:2015ogb,Low:2017mlh,Campiglia:2017dpg,Hamada:2017atr,Francia:2018jtb,Rodina:2018pcb,Distler:2018rwu}.    } \cite{Strominger:2013jfa,Strominger:2014pwa,He:2014cra,He:2014laa,Kapec:2015ena,He:2015zea,Campiglia:2015qka,Pasterski:2015tva,Pasterski:2015zua}.  Those works established a template for relating soft theorems to asymptotic symmetries.  We will follow that template closely in this work.  Since the underlying template is the same, let us just comment on the differences.

First, our symmetry is Geroch symmetry.  This is a global symmetry of the sigma model.  Thus, these are not just asymptotic symmetries, they are really ``everywhere symmetries.''  Of course, they also act asymptotically and we will phrase our discussion in those terms so that we can see the connection to infrared triangles.  

Second, let us clarify that the Geroch group is infinite dimensional but we are only ever going to use two Geroch transformations in this paper. 
One of the Geroch transformations we consider is a shift symmetry.  The other is a slightly more complicated version of a shift symmetry which reduces to a shift symmetry in the free theory. 
In the future, it would be very interesting to work out the full set of $\SS$-matrix constraints coming from Geroch symmetry.   

So far we have discussed soft theorems and asymptotic symmetries.  The third corner of the infrared triangle is the memory effect.  However, since cylindrical gravitational waves have a zero instead of a pole, there is no memory effect.  In other words, there is no dynamical way to cause a transition between Geroch-related vacua.  This stands in stark contrast to the memory effect for ordinary gravitons.  In the case of ordinary gravitons, hard radiation can cause transitions between supertranslation-related vacua.

In sum, the fact that cylindrical gravitational waves have a soft zero instead of a soft pole is a manifestation of Geroch symmetry and of the extraordinary simplicity of cylindrical gravitational waves.  This simplicity would seem to make cylindrical gravitational waves an attractive target for holography.

\section{Geroch symmetry}

Cylindrical gravitational waves are effectively two dimensional because we can ignore the $z$ and $\phi$ directions. 
Instead of general relativity, we can use a dimensional reduction of general relativity to two spacetime dimensions.  
The reduced theory is a nonlinear sigma model.  
The 2d spacetime metric is
\beq
ds^2 = -dt^2 + dr^2 \,.
\eeq
Space is a half-line ($r\geq0$).  
Future infinity, $\Sigma^+$, is the line at $t=\infty$ with coordinate $r$.  
$\Sigma^+$ has boundaries at $r=0$ and $r=\infty$ which we denote $\Sigma^+_0$ and $\Sigma^+_+$.

The equations of motion of the sigma model are \cite{Breitenlohner:1986um,Nicolai:1991tt,Schwarz:1995td,Schwarz:1995zy,Lu:2007zv}
\begin{align}
- \partial_a \left(r \partial^a X^1 \right) + r e^{2 X^1} \partial_a X^2 \partial^a X^2 &= 0 \,,	\label{eq:eom1} \\
\partial_a \left( r e^{2 X^1} \partial^a X^2 \right) &= 0 \,.		\label{eq:eom2}
\end{align}
To lighten the notation, we have set the coupling constant $g=1$.  
Here $X^\mu = X^\mu(t,r)$ is a pair of scalar fields ($\mu=1,2$).  
They encode the two physical polarizations of the cylindrical gravitational wave.  
Eqs. \eqref{eq:eom1}--\eqref{eq:eom2} have a pair of global symmetries:
\beq
\delta X^1 = -1 \,, \qquad
\delta X^2 = X^2 \,,			\label{eq:shift1}
\eeq
and
\beq
\delta X^2 = -1 \,.				\label{eq:shift2}
\eeq
Eq. \eqref{eq:shift2} is a shift symmetry.
Eq. \eqref{eq:shift1} is a shift symmetry in the free theory and something slightly more complicated in the interacting theory.  
This sigma model actually has an infinite dimensional global symmetry called the Geroch group.  
Eqs. \eqref{eq:shift1} and \eqref{eq:shift2} are the simplest Geroch transformations.  Most of the more complicated Geroch transformations are ``hidden symmetries'' in the sense that they involve nonlocal transformations of the fields.

The sigma model reduces to a free theory on $\Sigma^+$.  
In the free theory, the Noether charges for eqs. \eqref{eq:shift1}--\eqref{eq:shift2}  are\footnote{These symmetries have also been studied by Niedermaier \cite{Niedermaier:2000ud}, who also concluded that they are spontaneously broken in the quantum theory. }
\beq\label{eq:Qp}
Q^+ = \int_{\Sigma^+} dr \, r \partial_t X \,.
\eeq
There is no ``hard charge'' on the rhs.  This means there is no memory effect for these charges.  This is quite different from the story for supertranslation charges.  Supertranslation charges receive contributions from hard and soft gravitons.

Define a dual field, $\hat{X}$, by
\begin{align}
r \partial_t X &= \partial_r \hat{X} \,,	\label{eq:dual1} \\
r \partial_r X &= \partial_t \hat{X} \,.	\label{eq:dual2}
\end{align}
The consistency conditions for $\hat{X}$ are just the free equations of motion, $\partial_a (r \partial^a X) =  0$.  Eqs. \eqref{eq:dual1} and \eqref{eq:dual2} only define $\hat{X}$ up to a constant.  We fix the constant by requiring 
$\hat{X}|_{\Sigma^+_0} = 0$. 
The Noether charge \eqref{eq:Qp} becomes
\beq\label{eq:Qpp}
Q^+ = \hat{X} \vert_{\Sigma^+_+} \,.
\eeq
This is the asymptotic Geroch charge at the ``celestial point.''

Define $\pi^\mu = r \partial_t X^\mu$.  
The non-vanishing commutators on $\Sigma^+$ are
\beq
\left[ \pi^\mu(r) , X^\nu(r') \right] = - i \delta^{\mu\nu} \delta(r - r') \,.
\eeq
Thus
\beq\label{eq:QX}
\left[ Q^{+\mu} , X^\nu \right] = - i \delta^{\mu\nu} \,.
\eeq
So $Q^+$ generates shifts of $X$, as it should.

Now consider the analogous structure at past infinity.  
Past infinity, $\Sigma^-$, is the line at $t=-\infty$ with coordinate $r$.   
$\Sigma^-$ has boundaries at $r=0$ and $r=\infty$ which we denote $\Sigma^-_0$ and $\Sigma^-_+$. 
As before, we require $\hat{X} \vert_{\Sigma^-_0} = 0$.  
The Geroch charges are
\beq
Q^- = \int_{\Sigma^-} dr \, r \partial_t X = \hat{X} \vert_{\Sigma^-_+} \,.
\eeq
For notational simplicity, we use the same notation for $X$ and $\hat{X}$ at $\Sigma^\pm$.

The shifts generated by $Q^\pm$ come from the global symmetries \eqref{eq:shift1}--\eqref{eq:shift2}.  
So it is natural to require that shifts at $\Sigma^+_+$ are matched by shifts at $\Sigma^-_+$.  
This is our version of the ``antipodal matching'' condition.  
Only here the ``celestial sphere'' is a celestial point ($\Sigma^\pm_+$), so the antipodal part of the matching condition is trivial.

\section{Soft theorem}

In this section we consider the consequences of Geroch symmetry for the semi-classical $\mathcal{S}$-matrix.  
Let us denote an in (out) state comprised of $n$ ($m$) particles with energies $k_i^{\rm in}$ ($k_i^{\rm out}$) on the celestial point by
$\ket \equiv | k_1^{\rm in} , \cdots , k_n^{\rm in} \rangle$ 
($\bra \equiv \langle k_1^{\rm out} , \cdots , k_n^{\rm out} |$) .  
The $\mathcal{S}$-matrix elements are then denoted by $\bra \mathcal{S} \ket$.  
The quantum version of the classical invariance of scattering under Geroch symmetry is
\beq\label{eq:conservationlaw}
\bra \left( Q^+ {\mathcal{S}} - \mathcal{S} Q^- \right) \ket = 0 \,.
\eeq
Defining 
\beq
Q = Q^+ - Q^- \,,
\eeq
and the time-ordered product
\beq
\normord{ Q \mathcal{S} } \, = Q^+ {\mathcal{S}} - \mathcal{S} Q^- \,,
\eeq
eq. \eqref{eq:conservationlaw} becomes
\beq\label{eq:ward}
\bra \normord{ Q \mathcal{S} } \ket = 0 \,.
\eeq

At late times $t\rightarrow \infty$, the field becomes free and can be approximated by the mode expansion
\beq\label{eq:modes}
X(t,r)
	= \frac{1}{\sqrt{2}} \int_0^\infty dk 
		\left[ a_+(k) e^{-ikt} + a_+(k)^\dag e^{ikt} \right] J_0(k r) \,.
\eeq
The creation and annihilation operators on $\Sigma^+$, $a_+^\dag$ and $a_+$, obey
\beq
\left[ a^\mu_+(k) , a^\nu_+(k')^\dag \right] = \delta^{\mu\nu} \delta(k - k') \,.
\eeq
The dual field is
\beq\label{eq:dualmodes}
\hat{X}(t,r)
	= - \frac{i}{\sqrt{2}} \int_0^\infty dk 
		\left[ a_+(k) e^{-ikt} - a_+(k)^\dag e^{ikt} \right] r J_1(k r) \,.
\eeq

Eq. \eqref{eq:modes} is an expansion in energy eigenmodes.
Energy eigenmodes do not have a well-defined limit at $r=\infty$.  
To evaluate the Geroch charges at $\Sigma^+_+$, it is better to use wavepackets.  A natural way to get wavepackets is to Laplace transform the energy eigenmodes.  Define
\begin{align}
A_\pm(s) 			&= \int_0^\infty dk \, e^{-sk} a_\pm(k) \,,		\label{eq:laplace1} \\
A_\pm(\bar{s})^\dag	&= \int_0^\infty dk \, e^{-s k} a_\pm(k)^\dag 	\label{eq:laplace2} \,.
\end{align}
The inverse Laplace transforms are
\begin{align}
a_\pm(k) 			&= \int_C \frac{ds}{2\pi i} \,
								e^{sk} A_\pm(s) \,,		\label{eq:inverse1} \\ 
a_\pm(k)^\dag 		&= \int_C \frac{ds}{2\pi i} \,
								e^{sk} A_\pm(\bar{s})^\dag \,.	\label{eq:inverse2}
\end{align}
Insert eqs. \eqref{eq:inverse1}--\eqref{eq:inverse2} into \eqref{eq:dualmodes} to get
\beq
\hat{X}(t,r)
	= - \frac{i}{\sqrt{2}} \int_C \frac{ds}{2\pi i} \int_0^\infty dk \, 
		e^{sk} \left[ A_+(s) e^{-ikt} - A_+(\bar{s})^\dag e^{ikt} \right] r J_1(k r) \,.
\eeq
Do the $k$ integral to get
\beq\label{eq:dualwavepackets}
\hat{X}(t,r)
	= - \frac{i}{\sqrt{2}} \int_C  \frac{ds}{2\pi i} 
		 \left[ A_+(s) \varphi_{-s}(t,r)  - A_+(\bar{s})^\dag \varphi_{-\bar{s}}(t,r)^*  \right]  .
\eeq
We have defined
\begin{align}
\varphi_s(t,r) 			&=  \int_0^\infty dk \, e^{-( s + it )k} r J_1(k r) \,, \\
\varphi_{\bar{s}}(t,r)^*	&=  \int_0^\infty dk \, e^{-( s - it ) k} r J_1(k r) \,.
\end{align}
Eq. \eqref{eq:dualwavepackets} is the wavepacket expansion of $\hat{X}$ at $\Sigma^+$.
Now observe $\varphi_s \vert_{\Sigma^+_+} = 1$.
Thus \eqref{eq:Qpp}
\beq\label{eq:Qppackets}
Q^+	= \hat{X} \vert_{\Sigma^+_+} 
	= - \frac{i}{\sqrt{2}}  \int_C  \frac{ds}{2\pi i} 
		 \left[ A_+(s)  - A_+(\bar{s})^\dag  \right]  .
\eeq
These are the Geroch charges at $\Sigma^+_+$ in $s$-space.

To go back to $k$-space, insert eqs. \eqref{eq:laplace1}--\eqref{eq:laplace2} into \eqref{eq:Qppackets}:
\beq
Q^+	= - \frac{i}{\sqrt{2}} \int_0^\infty dk \,  \int_C  \frac{ds}{2\pi i}  \, 
		 e^{-s k} \left[ a_+(k)  -  a_+(k)^\dag  \right]  .
\eeq
Do the $s$ integral to get
\beq
Q^+	= - \frac{i}{\sqrt{2}} \int_0^\infty dk \,  \delta(k) 
		\left[ a_+(k)  -  a_+(k)^\dag  \right]  .
\eeq
Do the $k$ integral to get
\beq\label{eq:Qpk}
Q^+	= - \frac{i}{\sqrt{2}}  \left[ a_+(0)  -  a_+(0)^\dag  \right]  .
\eeq
A parallel construction at $\Sigma^-_+$ gives
\beq\label{eq:Qmk}
Q^-	= - \frac{i}{\sqrt{2}}  \left[ a_-(0)  -  a_-(0)^\dag  \right]  .
\eeq
Eqs. \eqref{eq:Qpk} and \eqref{eq:Qmk} are the Geroch charges at $\Sigma^\pm_+$.

Recall
\beq
Q = Q^+ - Q^- \,.
\eeq
The Ward identity \eqref{eq:ward} becomes
\beq
\bra \normord{ (Q^+ - Q^-) \mathcal{S} } \ket = 0 \,.
\eeq
Using \eqref{eq:Qpk} and \eqref{eq:Qmk}, this equation becomes
\beq\label{eq:softtheorem}
\bra a_+(0) \mathcal{S} \ket 
	= \bra \mathcal{S}  a_-(0)^\dag \ket = 0 \,.
\eeq
We have used\footnote{We do not have a general proof of this identity but it is true in the examples we have checked.  In particular, it is true for the amplitudes computed in our earlier work \cite{penna2024a}.}
\beq\label{eq:crossing}
\bra \mathcal{S}  a_-(0)^\dag \ket
	= - \bra a_+(0) \mathcal{S} \ket  \,.
\eeq
Eq. \eqref{eq:softtheorem} is the soft zero for cylindrical gravitational waves.  

In a recent paper \cite{penna2024a}, we computed the 2-particle tree-level $\SS$-matrix for cylindrical gravitational waves.  It is easily checked that eq. \eqref{eq:softtheorem} is valid for those amplitudes.  This lends some concrete support to the soft theorem \eqref{eq:softtheorem}.

\subsection*{Acknowledgements}

I am grateful to Miguel Campiglia for comments on the manuscript.

\bibliographystyle{JHEP}
\bibliography{citegeroch,citestrominger,misc}

\end{document}